\documentclass[aps,prl,twocolumn,twoside,floatfix,amsmath,showpacs,footinbib]{revtex4-1}

\usepackage{graphicx}
\usepackage{natbib}
\usepackage{color}
\usepackage[normalem]{ulem}

\begin{document}

\newcommand{\pdcoo}{PdCoO$_2$}
\newcommand{\ptcoo}{PtCoO$_2$}
\newcommand{\pdcro}{PdCrO$_2$}
\newcommand{\coo}{CoO$_2$}
\newcommand{\naxcoo}{Na$_x$CoO$_2$}
\newcommand{\urusi}{URu$_2$Si$_2$}
\newcommand{\nancoo}[1]{Na\ensuremath{_{#1}}CoO$_2$}
\newcommand{\prbacuo}{PrBa$_2$Cu$_4$O$_8$}
\newcommand{\cdi}{CdI$_2$}
\newcommand{\tis}{TiS$_2$}
\newcommand{\mos}{MoS$_2$}
\newcommand{\wse}{WSe$_2$}
\newcommand{\nbse}{NbSe$_2$}
\newcommand{\amo}{AMO$_2$}
\newcommand{\twoonefour}{Sr$_2$RuO$_4$}

\newcommand{\mum}[1]{\ensuremath{#1\,\mu\textnormal{m}}}
\newcommand{\sovert}[1]{\ensuremath{#1\,\mu\textnormal{V K}^{-2}}}
\newcommand{\seebeck}[1]{\ensuremath{#1\,\mu\textnormal{V K}^{-1}}}
\newcommand{\resist}[1]{\ensuremath{#1\,\mu\Omega\,\textnormal{cm}}}
\newcommand{\mJmolK}[1]{\ensuremath{#1\,\textnormal{mJ mol}^{-1} \textnormal{K}^{-2}}}
\newcommand{\lorenzunits}{\ensuremath{\,\textnormal{W \Omega K}^{-2}}}
\newcommand{\wkm}[1]{\ensuremath{#1\,\textnormal{W K}^{-1}\textnormal{m}^{-1}}}
\newcommand{\dos}{DOS}
\newcommand{\gfac}{\ensuremath{g_{\textnormal{eff}}}}

\newcommand{\ie}{{\em i.e.}}
\newcommand{\eg}{{\em e.g.}}

\newcommand{\replace}[2]{\sout{#1} \textcolor{red}{#2}}
\newcommand{\tred}[1]{\textcolor{red}{#1}}
\newcommand{\tblue}[1]{\textcolor{blue}{#1}}

\title{Impact of short-range order on transport properties of the two-dimensional metal \pdcro{}}

\author{Ramzy~Daou}
\affiliation{Laboratoire CRISMAT UMR6508, CNRS / ENSICAEN / UCBN, 6 Boulevard du Mar\'echal Juin, F-14050 Caen, France}
\author{Raymond~Fr\'esard}
\affiliation{Laboratoire CRISMAT UMR6508, CNRS / ENSICAEN / UCBN, 6 Boulevard du Mar\'echal Juin, F-14050 Caen, France}
\author{Sylvie~H\'ebert}
\affiliation{Laboratoire CRISMAT UMR6508, CNRS / ENSICAEN / UCBN, 6 Boulevard du Mar\'echal Juin, F-14050 Caen, France}
\author{Antoine~Maignan}
\affiliation{Laboratoire CRISMAT UMR6508, CNRS / ENSICAEN / UCBN, 6 Boulevard du Mar\'echal Juin, F-14050 Caen, France}

\date{\today}


\begin{abstract}
We study the Hall and Nernst effects across the antiferromagnetic transition that reconstructs the quasi-2D Fermi surface of the metallic local moment antiferromagnet \pdcro{}. We show that non-monotonic temperature dependence in the Hall effect and a sign change in the Nernst effect above the ordering temperature cannot be understood within a simple single-band transport model. The inclusion of coherent scattering by critical antiferromagnetic fluctuations can qualitatively account for these features in the transport coefficients. We discuss the implications of this for the pseudogap phase of the cuprate superconductors, which have a similar Fermi surface and where the same transport signatures are observed.
\end{abstract}

\maketitle

The precursor phase to a continuous phase transition generically consists of a regime of critical fluctuations of the order parameter, the spectrum of which is controlled only by the broken symmetry associated with the ordered phase \cite{wilson1983}. 
Superconductivity, for example, is preceded by paraconductivity, where superconducting fluctuations contribute to the conductivity even above the transition temperature. Characteristics of the superconducting state such as the correlation length can be extracted from the paraconducting regime \cite{Aslamazov1968,Maki1989}. In general, however, the impact of symmetry breaking on transport properties depends on the details of the electronic structure. 

One view of the pseudogap phase of the cuprate superconductors is that it is the precursor to some as-yet unidentified order. This view is supported by the quantum oscillation \cite{doiron2007}, Hall \cite{boeuf2007} and Nernst \cite{cyr2009,daou2010} effect measurements which indicate that the Fermi surface is reconstructed, at least when superconductivity is suppressed by  strong magnetic fields. The fact that the transport properties change without apparent discontinuities makes the identification of the ordered state difficult. Short-range charge order has, however, been detected in the pseudogap phase \cite{wu2011,boeuf2013,blanco2014}, and this should have an impact on transport properties \cite{fresard1990}. 

Here we study the model metallic single-band local moment antiferromagnet \pdcro{}, and show that the critical fluctuations of the antiferromagnetic precursor phase modify the transport properties in a way that foreshadows the broken symmetry phase. The Hall and Nernst effect prove particularly sensitive to the precursor phase, with non-monotonic temperature dependence and unexpected sign changes. Including coherent magnetic scattering in a semi-classical transport calculation allows us to model these essential features. The simple quasi-2D Fermi surface makes this system a useful non-superconducting analogue to the cuprates. Comparison with the isostructrual non-magnetic material \pdcoo{} is also instructive.

The delafossite oxide \pdcro{} is a frustrated triangular-lattice antiferromagnet. It is unusual in that it combines highly anisotropic, very metallic conduction with local moment magnetism \cite{takatsu2009}. Triangular Pd layers giving rise to a simple single quasi-2D Fermi surface sheet are separated by layers of edge-sharing CrO$_2$ octahedra. The Cr$^{3+}$ moments order in a 120$^\circ$ planar antiferromagnetic configuration below $T_N=37.5$\,K \cite{takatsu2014}, which results in a reconstruction of the single band into an ambipolar Fermi surface containing electrons and holes. This reconstruction is observed via photoemission above and below $T_N$ \cite{sobota2013,noh2014}, and the reconstructed bands are seen at low temperature by quantum oscillation measurements \cite{ok2013,hicks2015}. The resistivity shows a small sharp drop at $T_N$, while the Hall effect becomes non-linear with magnetic field below $T_N$ \cite{takatsu2010b,ok2013}, a sign of multi-band conduction in a clean metal. The magnetic contribution to the specific heat is dominated by critical fluctuations near $T_N$ with a critical exponent $\alpha = 0.13$ that holds up to 60\,K\cite{takatsu2009}, a wide range compared to non-frustrated magnets.

Single crystal samples of \pdcro{} were grown by the metathetical reaction \cite{shannon1971,takatsu2010a}. Very thin single crystal platelets of typical size $0.5\times 0.5 \times 0.002$\,mm$^3$ were obtained. 
Contacts for transport measurements were made with 25\,$\mu$m gold wire using Dupont 6838 silver epoxy. 
Resistivity and Hall effect were measured using a standard 4-probe technique with current and field reversal. The samples are thin enough that the measurement is insensitive to out-of-plane transport. Thermal and thermoelectric transport were measured in situ using the same contacts using a one-heater, two-thermometer steady-state technique on a custom built sample holder. Magnetic field was applied along the c-axis.

\begin{figure}
\includegraphics{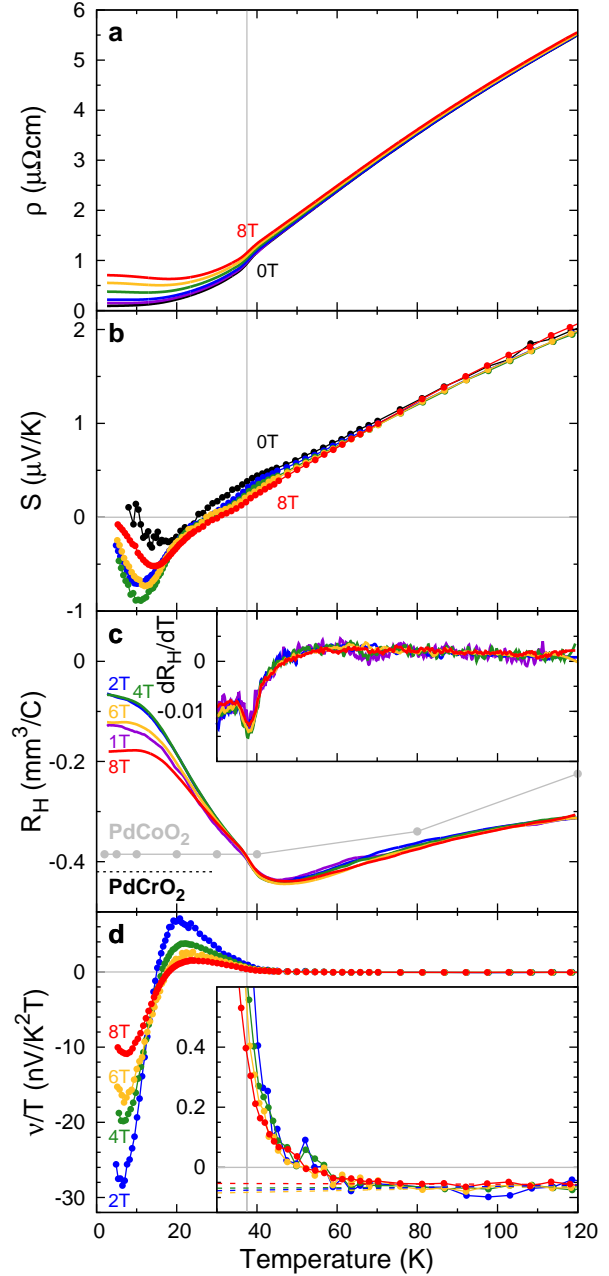}
\caption{\scriptsize{Magnetotransport properties of \pdcro{} in fields of 0\,T (black), 1\,T (purple), 2\,T (blue), 4\,T (green), 6\,T (yellow) and 8\,T (red) applied parallel to the c-axis.
a) A drop in resistivity occurs at $T_N$ (vertical grey line). $T_N$ is unaffected by these fields. 
b) The Seebeck coefficient for the same magnetic fields shows no pronounced feature at $T_N$ and a significant field dependence in the two-band regime below $T_N$.
c) $R_H$ decreases smoothly from 120-60\,K, in a similar fashion to the non-magnetic analogue \pdcoo{} (grey points from Ref.~\cite{takatsu2010b}). At 45\,K there is a minimum in $R_H$ and it begins to rise again. Below $T_N$ there is strong field dependence indicating the presence of two bands. The dashed black line indicates the expected value of $R_H$ for a single band, based on the value for \pdcoo{} scaled by the difference in lattice parameter. The inset shows the temperature derivative of $R_H$ which has a pronounced minimum at $T_N$. This also show the deviation from the high temperature behaviour begins around 50--60\,K.
d) The Nernst coefficient $\nu/T = N/TB$ is small and positive at high temperature. Below $T_N$ it becomes large, ambipolar and strongly field-dependent. The inset shows that the rise of $\nu/T$ from a high-temperature baseline (dashed lines) is clear around 60--70\,K.}
}
\label{fig:nernst}
\end{figure}

The resistivity $\rho$ (Fig.~\ref{fig:nernst}a) is similar to previous reports \cite{takatsu2010a,hicks2015}, with a pronounced drop just at $T_N$.
This is understood as a loss of scattering at the hotspots as gaps induced by magnetic ordering open. A positive magnetoresistance is visible everywhere, which becomes large at low temperatures. The residual resistivity is \resist{0.05}. The Seebeck coefficient $S$ (Fig.~\ref{fig:nernst}b) is very similar to the isostructural non-magnetic compound \pdcoo{} \cite{daou2015}, which has a very similar electronic structure, but there is little change on passing through $T_N$. This is reasonable as $S$ is expected to depend less directly on the scattering rate than $\rho$.


In Fig.~\ref{fig:nernst}c we show Hall effect measurements as a function of temperature for several magnetic fields. The Hall coefficient $R_H$ also has a change in slope at $T_N$, more clearly visible in the derivative. The change from a single- to a multi-band carrier spectrum qualitatively explains the large scale changes in $\rho$ and $R_H$, including the strong magnetic field dependence in the multi-band regime \cite{ok2013}. Close to $T_N$, the impact of critical fluctuations on transport is less easy to discern. The derivative $dR_H/dT$ suggests that the limiting high temperature behaviour starts to change at around 60\,K, between which and $T_N$ there is a minimum in $R_H$. In the non-magnetic analogue \pdcoo{}, $R_H$ has a similar temperature dependence above 40\,K (also shown in Fig.~\ref{fig:nernst}c), but becomes constant below this temperature as there is no reconstruction of the Fermi surface. The resistivity meanwhile has a sub-linear temperature dependence from $T_N$ up to room temperature that has been interpreted as indicating that magnetic excitations are active over a very wide temperature range \cite{hicks2015}. For comparison, in \pdcoo{} there is an anomalously super-linear temperature dependence that was explained by scattering of electrons from high-frequency optical phonons \cite{takatsu2007}.

The Nernst effect (Fig.~\ref{fig:nernst}d) is particularly sensitive to both fluctuations and reconstruction of the Fermi surface. The Nernst coefficient $\nu$ is defined by the transverse electric field $E_y$ generated by a longitudinal thermal gradient $\nabla_xT$ in the presence of a perpendicular magnetic field $B_z$, such that $\nu = E_y/B_z\nabla_xT$ (we use the vortex sign convention \cite{behnia2009}). For a single band the expectation is that it is constant with magnetic field and nearly zero, a result of 'Sondheimer cancellation'. In \pdcro{} for $T>>T_N$, the magnitude $0.07$nV/K$^2$T is close to the estimation, based on the Mott formula for the thermoelectric coefficients\cite{behnia2009}, that $\frac{\nu}{T} = \frac{\pi^2k_B^2}{3e} \frac{\partial\mu_H}{\partial\varepsilon}\Bigr|_{\varepsilon_F} \approx \frac{\pi^2k_B^2}{3e} \frac{\mu_H}{\varepsilon_F}$. $\mu_H$ is the Hall mobility from measurements, which gives $\nu/T = 0.2$nV/K$^2$T at 100K, while $\varepsilon_F = 0.8$\,eV is the bandwidth taken from electronic structure calculations.

In a multi-band system, $\nu$ need not be small as Sondheimer cancellation is suppressed by the presence of multiple carrier types. The mobility can be expected to have a complex energy dependence, resulting in a Nernst coefficient that is not linear in $B_z$, and with a strong temperature dependence \cite{bel2003}. This is consistent with the data for $T<T_N$. Magnetic breakdown may also play a role, as was postulated to explain the low temperature Hall effect \cite{ok2013}; the unconventional anomalous Hall effect has also been invoked to explain the low temperature Hall coefficient at $T<20K$ \cite{takatsu2010b}, but we would not expect this to be a factor above $T_N$.

In the range $T_N<T<60K$, the inset to Fig.~\ref{fig:nernst}d shows that $\nu/T$ begins to change from its limiting high temperature behaviour, changes sign, and starts to acquire a weak dependence on $B_z$ close to $T_N$. We interpret this as the effect of critical fluctuations associated with the antiferromagnetic transition on the Nernst signal. These enter as another scattering channel in addition to the electron-phonon scattering that typically dominates conduction in metals. The dependence of $\nu/T$ on magnetic field close to $T_N$ is further evidence of a magnetic origin of this contribution in this temperature range. 

A similar observation has been made in the case of LaFeAsO$_{1-x}$F$_x$.
At $x=0$, the spin-density-wave transition at 140\,K causes reconstruction of the Fermi surface and a coincident increase in magnitude of the Nernst signal \cite{kondrat2011}. At $x=0.05$, just beyond the region where order is stabilized, the Nernst signal behaves similarly although there is no static order \cite{hess2009}. Spin fluctuations associated with proximity to the order have been invoked to explain this anomalous behaviour. They are also thought mediate the pairing of the superconducting state. The non-magnetic Fermi surface consists of multiple sheets already, which does not rule out the possibility that complex temperature dependence of the normal-state Nernst effect may arise from the ambipolar mechanism. 

The cuprate superconductors have a single-band quasi-2D Fermi surface contour that is more closely analagous to \pdcro{}, and a similar Nernst response on entering the pseudogap phase. The normal state value of $\nu/T$ is small and constant at high temperature, and changes sign and increases in magnitude as the pseudogap phase is entered \cite{cyr2009,daou2010}. No order parameter has been conclusively identified for the pseudogap phase, although recent experiments show that short-range charge order is stabilised over some range of doping and temperature \cite{wu2011,boeuf2013,blanco2014}. The Nernst response of \pdcro{} to the short-range order of the precursor phase is a strong indicator that the very similar response in the cuprates is also due to the presence of a short-range ordered phase. The origin of this short range order is of course quite different. It is based on local moment magnetism isolated from the conducting planes in \pdcro{}, while it appears to arise from charge-density-wave instabilities within the copper oxide planes in the case of the cuprates. Although the Fermi surface is similar, the dispersion relations close to the Fermi energy need not be identical as a result, as indicated by the enhanced effective mass in overdoped Tl-2201 \cite{vignolle2008} ($5m_e$ compared to $1.5m_e$ here).

The question remains, in \pdcro{} (as well as in the cuprate superconductors), as to how precisely $\nu/T$ and $R_H$ can acquire non-monotonic temperature dependence or even change sign when there is only a single band in the electronic structure. To address this, we perform semiclassical transport calculations which show that increased diffusive scattering as $T\rightarrow T_N^+$ is not a sufficient explanation; in fact this would always have a tendency to increase the resistivity and cannot cause $\nu/T$ to change sign for a given bandstructure for any reasonable choice of relaxation time.

We use the electronic bandstructure of \pdcoo{} (obtained using Quantum Espresso and in good agreement with other calculations \cite{eyert2008,ong2010a}) as the input to calculations of the transport coefficients as the temperature approaches $T_N$. Spectroscopic probes of the bandstructure \cite{noh2014,sobota2013,ok2013,hicks2015} suggest that the bandstructure of \pdcro{} at $T>T_N$ is very similar to that of the non-magnetic analogue \pdcoo{}, scaled slightly to account for the difference in lattice parameter. The application of a finite $U=5$\,eV on the Co site was found to produce a Fermi surface in better agreement with dHvA measurements\cite{hicks2012} and so we present results for both $U=0$ and $U=5$\,eV. 

A previous transport calculation using the BoltzTraP package showed high anisotropy in the conductivity and Seebeck coefficient \cite{ong2012}, in good qualitative agreement with experimental results. This package cannot model the Hall and Nernst effects at strong magnetic fields, and so we calculate the transport integrals on this Fermi surface using \cite{ashcroftmermin}:
\begin{equation}
\boldsymbol{\sigma}^{(\alpha)} = e^2 \int\frac{d\mathbf{k}}{4\pi^3} (-\frac{\partial f}{\partial \varepsilon})
\tau(\varepsilon(\mathbf{k})) \mathbf{v(k)}\mathbf{\bar{v}(k)}(\varepsilon(\mathbf{k})-\varepsilon_F)^\alpha
 \label{eqn:transport1}
\end{equation}
where $\tau(\varepsilon(\mathbf{k}))$ is the relaxation time. The bandstructure enters through the momentum-dependent velocity. $\mathbf{v(k)}$ is the velocity at $\mathbf{k}$ and $\mathbf{\bar{v}(k)}$ is the weighted historical average velocity of all particles arriving at $\mathbf{k}$ without scattering. These particles are driven to precess around isoenergetic contours by the semiclassical magnetic field,
\begin{equation}
\mathbf{\bar{v}(k)} = \int_{-\infty}^0 \frac{dt}{\tau(\mathbf{k}(t))}e^{t/\tau(\mathbf{k}(t))} \mathbf{v}(\mathbf{k}(t))
 \label{eqn:transport2}
\end{equation}
where the only free parameter is $\tau(\varepsilon(\mathbf{k}))$. 

\begin{figure}
 \includegraphics{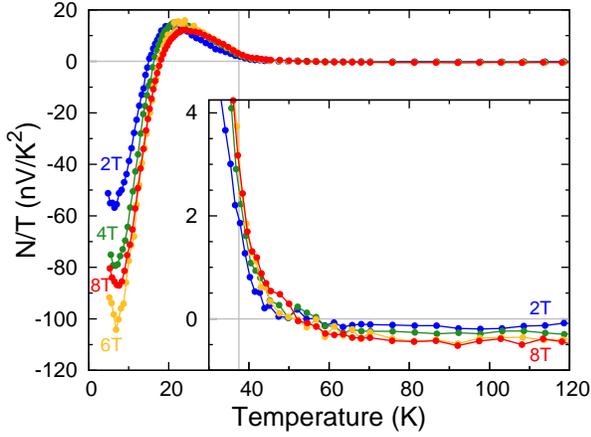}
\caption{Nernst signal $N/T$ as a function of temperature in magnetic fields of 2, 4, 6 and 8\,T.}
\label{fig:nernst2}
\end{figure}

The conductivity is $\boldsymbol{\sigma} = \boldsymbol{\sigma}^{(0)}$ while the thermoelectric tensor is $\boldsymbol{\alpha} = \boldsymbol{\sigma}^{(1)}$. 
We evaluate the integrals for $B || z$ at 600000 $k$-points near the Fermi energy. We calculated the Hall coefficient $R_H = (1/B_z) \sigma_{xy}/(\sigma_{xx}^2+\sigma_{xy}^2)$ and Nernst signal $N = (\alpha_{xy}\sigma_{xx} - \alpha_{xx}\sigma_{xy})/(\sigma_{xx}^2+\sigma_{xy}^2)$ as functions of $\omega_c\tau = eB_z\tau/m^*$ for magnetic fields applied along $z$. For comparison, the experimental values of $\omega_c\tau$ range from 0.01 (120\,K, 2\,T) to 0.4 (40\,K, 8\,T). The Nernst data plotted as $N/T$ rather than $\nu/T$ is shown in Fig.~\ref{fig:nernst2} since this calculated quantity is a universal function of $\omega_c\tau$.

\begin{figure}
 \includegraphics{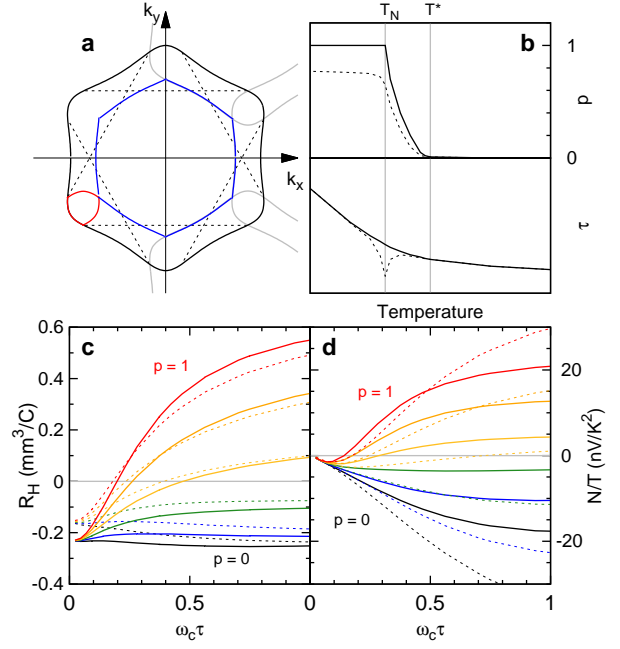}
\caption{Calculated transport coefficients. a) Cross section of Fermi surface of \pdcoo{} at $k_z=0$. The maximally warped parts of the Fermi surface lie at the corners of the pseudo-hexagonal cylinder. Gaps open at $T_N$ where the dashed lines intersect the FS, determined geometrically by the overlap of displaced copies of the FS by the $\sqrt{3}\times\sqrt{3}$ ordering wavevector (grey lines). Coherent magnetic scattering can occur between the hotspots connected by dashed lines with a probability $p$. When $p=1$, the reconstruction leads to electron (red) and hole orbits (blue). b) Schematic evolution of $p$ (top) and $\tau$ (bottom) with temperature. Magnetic breakdown acts to reduce $p$ (dashed line). Close to $T_N$ magnon-phonon scattering may reduce $\tau$ (dashed line). Transport properties begin to be influenced by short-range order at $T^*$. c) The calculated Hall coefficient $R_H$ as a function of $\omega_c\tau$ for different values of $p$ in steps of 0.2. The solid lines are for $U=5$\,eV and the dashed lines for $U=0$\,eV. d) The same for the Nernst signal divided by the temperature. The Nernst signal is much more sensitive than $R_H$ to the value of $U$. Note that $N/T = \nu B/T$ has been presented here since it is a function of $\omega_c\tau$, while the data of Fig.~\ref{fig:nernst}d are plotted as $\nu/T$ to highlight the linear regime of $N$ with magnetic field.}
\label{fig:model}
\end{figure}

These calculations show that the signs of the calculated transport coefficients $R_H$ and $N/T$ (also $S$) are fixed and cannot be changed by varying $\tau$ (see the curves for $p=0$ in Fig.~\ref{fig:model}). They depend only on the bandstructure around $\varepsilon_F$. It is impossible to explain the observed sign change in $\nu/T$ and the non-monotonic dependence of $R_H$ as $T \rightarrow T_N^+$. Introducing anisotropy in $\tau(\mathbf{k})$ does not change this.

We therefore extend the model to include magnetic scattering. Since the effect of incoherent magnetic scattering (such that the electron direction is randomised after collision) can only be to change $\tau(\mathbf{k})$, this mechanism cannot drive a sign change in the transport coefficients.

Coherent magnetic scattering does not randomise the electron velocity, but links two points on the Fermi surface via the magnetic wave-vector, as shown in Figure~\ref{fig:model}a. At $T_N$ this leads to the opening of gaps at these points and reconstruction of the Fermi surface. The opening of gaps also leads to a loss of incoherent magnetic scattering from these regions, and so a small drop in resistivity at $T_N$ in zero field. Above $T_N$ critical magnetic fluctuations exist on short length- and time-scales, and can cause scattering of both types without reconstructing the Fermi surface.

We model the effect of coherent magnetic scattering above $T_N$ by assigning a finite probability $p$ that electrons passing through a hotspot are scattered to the linked hotspot, and include this process in the evaluation of Eqn.~\ref{eqn:transport2}. That is, the historical paths of electrons arriving at $\mathbf{k}$ include several branching routes weighted by probabilites $p$ or $1-p$. Figure~\ref{fig:model} shows the result of this calculation as a function of $\omega_c\tau$ for the Fermi surfaces obtained with $U=0$ and 5\,eV for various values of $p$. While the magnetism is three dimensional, band folding in the $k_z$ direction will have little impact on the in-plane transport coefficients and has not been considered. The schematic evolution of $p$ and $\tau$ as a function of temperature are shown in Fig~\ref{fig:model}b. The purpose of the model is to ascertain whether behaviour qualitatively similar to that seen experimentally can be produced for reasonable values of $p$ and $\omega_c\tau$.

Sign changes in both $R_H$ and $N$ occur if $p$ and $\omega_c\tau$ are large enough. The sign changes do not need to occur at the same value of $\omega_c\tau$. There is a range of $p$ values close to 0.5 where a sign change is present in $N$ but not $R_H$, as is seen in experiments. The comparison between $U=0$ and $U=5$\,eV shows that this kind of behaviour depends somewhat on the details of the bandstructure. As expected, $R_H$ is less sensitive to $U$, since in the limit of high $\omega_c\tau$ it converges on the expected Drude value in both cases. $N$ is more sensitive to the energy dependence of the bandstructure, which is quite different in the two cases even though the volume enclosed by the Fermi contour is the same. The less warped Fermi surface with $U=5$\,eV therefore results in a smaller $|N|$. A further quantitative difference is that $N$ changes sign at lower $\omega_c\tau$ than $R_H$ for $U=5$\,eV, while the reverse is true for $U=0$. Further fine tuning of the sign changes may be achieved by introducing anisotropic scattering $\tau(\mathbf{k})$, which would allow the relative contribution of the electron-like paths (formed from the corners of the FS) and the hole-like paths to be controlled.

One clear weakness is that the calculation does not reproduce well the apparent insensitivity to magnetic fields (at least up to 8\,T) of $R_H$ and $\nu/T$ that is seen experimentally even during the sign change in $\nu/T$. It also implies that a sign change in $R_H$ is a likely consequence of the reconstruction of the Fermi surface, which is not the case experimentally. This may be the result of magnetic breakdown, which reduces $p$ as a function of $\omega_c\tau$.


These calculations suggest that for reasonable values of $\omega_c\tau$, qualitative features such as non-monotonic temperature dependence or sign changes in single-band transport coefficients require an ingredient beyond ordinary diffusive scattering mechanisms. This is provided by short-lived, short-range fluctuations of the order parameter in the precursor phase above the phase transition. In the case of \pdcro{}, the result is the foreshadowing of the transport properties in the ordered phase, as the fluctuations temporarily break the same symmetry as the order parameter.

In the case of the cuprates only short-range order has been identified, but the transport coefficients smoothly connect the low-field and high-field states where low frequency quantum oscillations indicate that gaps have opened \cite{chang2010}. It is not clear whether this short-range order possesses the correct symmetry that would lead to the observed quantum oscillation spectrum in the long-range ordered limit, given that they are only observed in high magnetic fields \cite{allais2014}. The changes in sign and magnitude of the Hall and Nernst effects in the cuprate superconductors are qualitatively very similar to those observed in \pdcro{} for $T>T_N$, suggesting that short-range order drives a reconstruction of the Fermi surface; whether that reconstruction is modified in the long-range limit is unclear. The qualitative features of the transport results seem to be insensitive to the microscopic nature of the interaction between the incipient order and the conduction electrons.

In conclusion, we have shown that all features of the Hall and Nernst response of the single-band quasi-2D metal \pdcro{} above the magnetic phase transition can be understood as the result of the associated short-range order coupling to the electronic states. The sign change in the Nernst effect in particular does not require long-range order, but does require that the order parameter associated with short range order would reconstruct the Fermi surface in the long-range limit.

\end{document}